\documentclass[pre,floatfix,twocolumn, 10pt]{revtex4-1} 
\usepackage{graphicx}
\usepackage{amssymb}
\usepackage{natbib}
\usepackage{dcolumn}
\usepackage{bm}
\usepackage{color}
\usepackage[colorlinks=true, linkcolor=blue, citecolor=blue]{hyperref}
\usepackage{subfigure}
\usepackage{graphicx,psfrag,xspace}
\usepackage{color}
\usepackage{amssymb,amsfonts,amsmath}
\usepackage{setspace}

\begin{document}

\author{E. A. Jagla} 
\affiliation{Comisi\'on Nacional de Energ\'{\i}a At\'omica, Instituto Balseiro (UNCu), and CONICET\\
Centro At\'omico Bariloche, (8400) Bariloche, Argentina}

\title{Elasto-plastic models of the yielding transition with stress-dependent transition rates}

\begin{abstract} 

Elasto-plastic models are among the most successful ways to study the critical properties of the plastic yielding transition
of amorphous solids. Typically these models are studied under a condition of constant transition rates from one plastic configuration to another, and in this form they predict the existence of well defined critical exponents that display universality, in the same sense that in standard equilibrium phase transitions. I show however that very naturally the transition rates must not be taken as a constant, but dependent of the local stress excess above the critical value. This modification in the model is seen to affect the values of some of the exponents of the transition, concretely, of the dynamical exponents that are related to the speed at which the system is driven. I argue about the reason for this dependence, claiming that it is due to the quasi-mean field nature of the plastic yielding transition originated in the fact that elastic interactions are long range.

\end{abstract}

\maketitle

\section{Introduction}

In recent years there have been great advances in the understanding of the yielding transition of amorphous solids. A wealth of both experimental and theoretical work has allowed to reach a good understanding of this kind of transitions that occur in a variety of materials of great technological importance \cite{bonn}. In the athermal case, in which temperature plays no fundamental role, the transition is known to be conducted by an interplay between the elasticity of the material and plastic effects that occur due to non-affine reaccommodations of small portions of the system \cite{ba,stz}. Although these plastic events are in principle uncorrelated because of the amorphous characteristics of the sample, the elastic interactions create correlations that favor the appearance of large avalanches of plastic rearrangements, which are one of the most typical properties of the transition. 

The theoretical understanding of the plastic depinning transition has benefited from concepts previously used to understand the depinning transition of elastic manifolds evolving on disordered substrates. In this case, a paradigm was constructed based on the analogy between this non-equilibrium, athermal transition and the classical theory of critical phenomena in temperature driven transitions\cite{fisher,kardar}. 

Elasto-plastic models (EPM) provide a description of plastic yielding transition that allows the best comparison with models for elastic depinning \cite{ferrero_rmp}. In EPM the evolution of stress as well as the plastic strain in the system is monitored. Under a condition of uniform and stationary load, stress increases uniformly across the system. When the local stress overpasses some critical value, the local plastic strain increases, causing a reduction of the local stress, and also a modification of the stress in every other point under the effect of an elastic propagation kernel known as the Eshelby propagator \cite{eshelby1,eshelby2}. The characteristics of the Eshelby kernel greatly determine the characteristics of the plastic yielding transition and the differences with its elastic depinning counterpart. The Eshelby kernel has a $\sim 1/r^d$ spatial decay ($d$ being the spatial dimension), and thus it is a long range interaction. Also, it has alternating signs depending of the direction, with a quadrupolar symmetry. This symmetry, and the zero-lines along particular directions are the responsible of the special avalanche correlations in the form of slip planes along easy directions that are observed at the yielding transition. 

There are slight variations among different versions of EPM found in the literature. However, most of them share the main fundamental ingredients.
The version I present here is closely related to the one discussed in Ref. \cite{pnas}. The model is defined in terms of the local stresses $\sigma_i$ and the local plastic deformations $\gamma_i^{pl}$. The average value of the local stresses defines the applied external stress $\sigma=\overline{\sigma_i}$. If the local stress is larger than a threshold value $\sigma_i^{th}$ then there is a probability per unit of time $\lambda$ that this site suffers a plastic rearrangement increasing its local plastic strain in some amount $\delta\gamma_i^{pl}$. The plastic strain increase produces an instantaneous modification of the stress in all sites mediated by the Eshelby interaction:

\begin{equation}
\delta\sigma_j=\sum_i G_{ji}\delta\gamma_i^{pl}
\label{eq1}
\end{equation}
The strain rate $\dot\gamma$ in the system is defined as the average rate of plastic deformation: $\dot\gamma=d\overline{\gamma_i^{pl}}/dt$.

\begin{figure}
\includegraphics[width=6cm,clip=true]{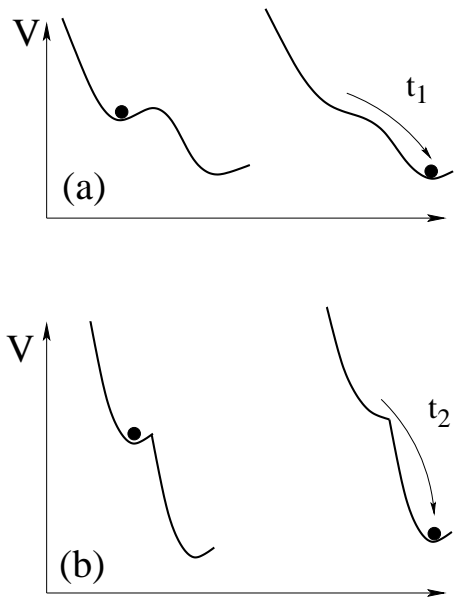}
\caption{Sketch of the potencial energy for the transition from a metastable state that is destabilizing to a new stable state. The transition time depends
on the potential being of the ``smooth" type (a), or the ``parabolic" type (b).
\label{dibu}
}
\end{figure}

The main point on which I want to focus is the rule that governs the increase of local plastic deformation. As it was stated, when $\sigma_i>\sigma_i^{th}$ there is a probability $\lambda$ per unit time that a plastic deformation occurs. In one form or another, in all versions of EPM presented so far, the value of $\lambda$ is taken as a constant, which essentially sets the time scale of the problem. However, some analysis reveals that this quantity should not be taken as constant. The plastic strain increase when the local stress threshold is overcome can be described formally as the passage between a local state that becomes unstable, 
to a new stable state, as soon as $\sigma_i>\sigma_i^{th}$. The situation is sketched in Fig. \ref{dibu}. Assuming a smooth form 
of the local potential energy on which the system evolves, the typical time $t_0$ needed to move to the new minimum is a function of the ``degree of instability" $\sigma_i-\sigma_i^{th}$, such that $t_0\sim (\sigma_i-\sigma_i^{th})^{-1/2}$. If we want to maintain an implementation in terms of transition rates, the conclusion is that the rate $\lambda$ is stress dependent, in such a way that $\lambda\sim (\sigma_i-\sigma_i^{th})^{1/2}$. Although the case of a smooth potential is the natural situation to be considered, it has to be mentioned that a situation of constant rate is obtained  if the potential is assumed to be composed of concatenated parabolic pieces. In this situation,  the time it takes for the system to reach the new minimum when $\sigma_i>\sigma_i^{th}$ is roughly constant, independent of the degree of instability. Then we can conclude that the prescription of a constant transition rate is compatible with a plastic potential formed by the concatenation of parabolic pieces. 

The main point of the present paper is to investigate the effect of the two different forms of the transition rate: constant (as in EPM studied so far) and $\sim (\sigma_i-\sigma_i^{th})^{1/2}$, which I argue is the most natural case to consider. 
I will refer to the two cases as ``uniform rates", and ``progressive rates".

Before going into the simulation details I want to discuss why this change of rates may be expected to have any effect at all in the critical properties of the transition. 
In the depinning counterpart, the consideration of different forms of the pinning potential was already made in early studies by Fisher (see in particular Ref. \cite{fisher2}). In those seminal works it was concluded  that the form of the local (microscopic) potential does not influence the critical (large scale) behavior of the system. These results were then confirmed by the use of 
the Functional Renormalization Group\cite{frg1,frg2,frg3}: in the large scale the potential effectively has a singular correlator, which is like saying that the potential looks as having cusps (as those of the concatenated parabola) even if the starting microscopic potential was smooth. The only exception to this situation is when the interaction is mean field: if all particles interact equally with all others, it is then straightforward to transform the model to a one-particle version that is seen to be equivalent to the Prandtl-Tomlinson model of friction\cite{ptmodel}. This model has a critical force for vanishingly small velocity, and the exponent of the force increase at small velocities depends in fact on the form of the potential.

Previous simulations in related versions of the plastic yielding model\cite{jagla2017}, and the present ones using different form of the rate law unambiguously reveal that the critical exponents associated with the dynamical evolution depend on the form of the rates. On the contrary the static exponents (for instance those associated with avalanche size distribution) are independent of the rates instead. In the light of the results for the depinning transition, these findings are a strong evidence that the plastic yielding transition is some kind of mean field transition.

\section{Simulation Protocols}

Simulations presented here are done in two dimensional systems, with two different protocols.

1-{\em Constant stress simulations}

In this case an initial set of values $\sigma_i$ is chosen such that its mean value matches the externally applied stress $\sigma$, i.e., $\overline {\sigma_i}=\sigma$. The simulation uses an elementary time step $\delta t$, and follows the following protocol:

(a) Unstable sites (those for which $\sigma_i>\sigma_i^{th}$) are destabilized with probability $p=\lambda \delta t$. For simplicity all thresholds are taken equal in the simulations: $\sigma_i^{th}\equiv 1$

(b) Let $j$ denote the destabilized sites determined at step (a). Then choose the values of plastic strain increases $\delta \gamma^{pl}_j$ from an exponential distribution with average value $\delta \gamma^{pl}_0$.

(c) Recalculate the stress across the system as:
$\sigma_i=\sigma_i+\delta\sigma_i$ with 
\begin{equation}
\delta\sigma_i=\sum_j G_{ij}\delta \gamma^{pl}_j
\end{equation}
where the sum is meant to be done over the destabilized sites.

(d) Go to step (a).

The form of $G_{ij}$ is obtained by Fourier inverting the expression
\begin{equation}
G_{\bf q}=\frac{(q_x^2-q_y^2)^2}{(q_x^2+q_y^2)^2}
\end{equation}
with $G_{{\bf q}=0}=0$, where $q_x^2$, $q_y^2$ must be understood in a square numerical mesh of size $L\times L$ as 
\begin{equation}
q_{x,y}^2\equiv 2-2\cos\left (\frac{\pi n_{x,y}}{L}\right )
\end{equation}
and $n_{x,y}=0,..., L-1$.

Note that $\sum_i G_{ij}\sim G_{{\bf q}=0}=0$, then step (c) preserves the average stress.
Constant stress simulations are used to determine $\dot\gamma$ in terms of $\sigma$. 
$\dot\gamma$ is calculated as the average fraction of unstable sites along the simulation.
In this way the flow exponent $\beta$ defined through  $\dot\gamma\sim (\sigma-\sigma_c)^\beta$ can be calculated.
Constant stress simulations reach a frozen state if there are no unstable sites at a given step. To access the vicinity of the critical point (with typically very few unstable sites), larger system sizes must be used to avoid this situation.

2-{\em Quasi-static simulations}

This form of the simulations is aimed to detect individual avalanches, and then to be able to collect statistics of size, duration, 
inter-events time, etc. This is a constant strain simulation, in which every plastic event is supposed to reduce the average stress in the system.
This is accomplished by modifying the form of the interaction kernel. Instead of the $G_{ij}$ of protocol 1, a $G'_{ij}\equiv G_{ij}-\kappa/N$ is used, with $N\equiv L\times L$ the total number of sites in the system, and $\kappa$ a fixed parameter (I will use $\kappa=3$ in the simulations below). The protocol then proceeds as follows.

Given a stable stress distribution $\sigma_i<1$, the average stress $\sigma\equiv\overline \sigma_i$ is increased until at one site $\sigma_i=1$. Now protocol 1 is followed using $G'_{ij}$ instead of $G_{ij}$. Since in this case every plastic strain increase $\delta\gamma^{pl}$ produces a reduction of the average stress by an amount $\delta\sigma=\kappa\delta\gamma^{pl}  /N$, the process is guaranteed to terminate. In this way an avalanche is generated, and its size $S$ and duration $T$ may be recorded.

Whatever of the two protocols are used, two possibilities for the rate $\lambda$ are considered: in the case of {\em uniform }rates, $\lambda$ is taken as an constant, in concrete $\lambda=1$ (different values of $\lambda$ affect only the time scale of the problem, and rescale directly the value of $\dot\gamma$ obtained). In the case of {\em progressive} rates $\lambda= (\sigma_i-1)^{1/2}$. Note that in general, a rate law depending on a continuous parameter $\eta$, such that $\lambda\sim (\sigma_i-1)^\eta$ could be studied. However I prefer to focus on the two physical cases, namely $\eta=0$ for potentials with cusps, and $\eta=1/2$ for smooth potentials.

\begin{figure}
\includegraphics[width=8cm,clip=true]{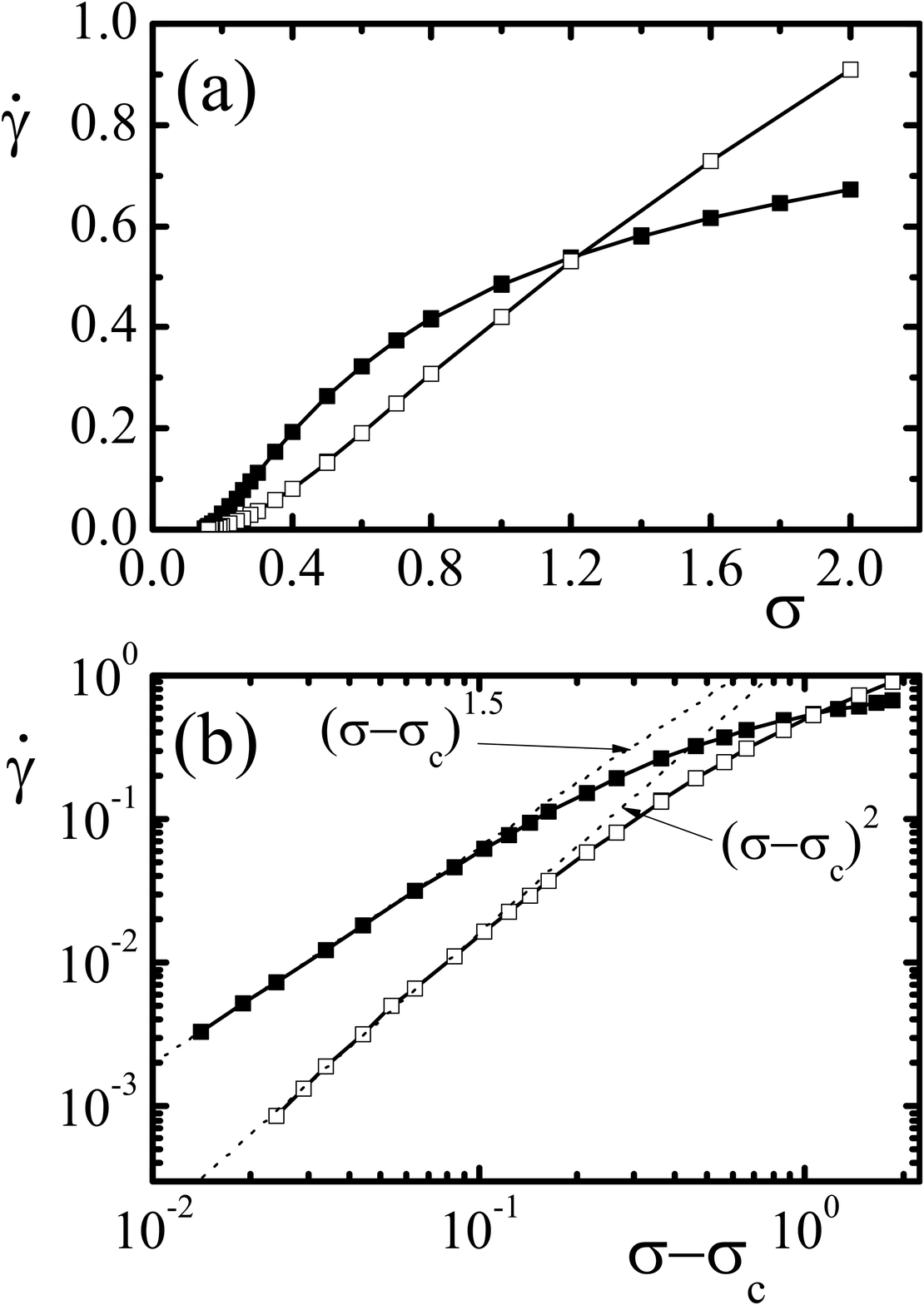}
\caption{Flow curves in a system of size 512$\times$ 512, for uniform (full symbols) and progressive (open symbols) transition rates. 
\label{beta}
}
\end{figure}

\section{Results}

The results presented in Fig. \ref{beta} correspond to constant stress simulations, and show the flow curves ($\dot\gamma$ vs. $\sigma$) obtained with the two forms of transition rates. The first point to be emphasized is that the critical stress $\sigma_c$ coincides for the two forms of the rates.
In fact, different forms of the rate only modify the time scale of the jumping process, but do not alter the stability of each configuration. In particular, a configuration is stable  independently of the form of the rates that are used when there are unstable sites. This is particularly important for the numerical simulations, as the value of $\sigma_c$ used in panel (b) of Fig. \ref{beta} is the same for both curves, and allows a more reliable comparison of the results for the two rates. The results clearly indicate that the values of $\beta$ are different, and are given by $\beta\simeq 1.5$ (uniform rates) and $\beta\simeq 2.0$ (progressive rates).

\begin{figure}
\includegraphics[width=8cm,clip=true]{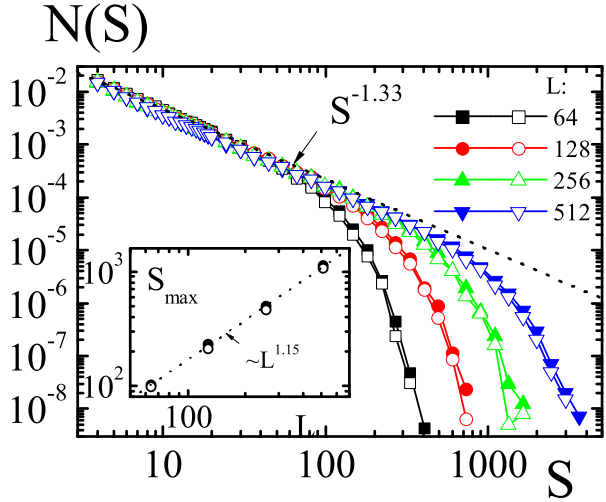}
\caption{Size distribution of avalanches for system of different sizes, using uniform (full symbols) and progressive (open symbols) rates. Inset: fitted values of $S_{max}$ as a function of system size for the two forms of the rates.
\label{ndes}
}
\end{figure}

Next, I show the results of numerical simulations performed with the quasi-static protocol. These simulations allow to generate a collection of avalanches to which different statistical properties can be measured.
The most important monitored quantities include: Avalanche size $S$, avalanche duration $T$, and stress increase $\delta \sigma$ to trigger the next avalanche. It has to be emphasized that in cases (like depinning) in which the kernel $G_{ij}$ is non-negative, there are theorems that guarantee that quantities characterizing the avalanche as a whole (like the previously mentioned quantities $S$, $T$, $\delta\sigma$) do not depend on the form of the rates (Middleton theorems\cite{middleton}). However, for yielding, the alternating sign property of $G_{ij}$ makes possible that some particular site is activated under a particular form of the transition rates, while it is not with some other form of the rates. Then the question of equivalence or not of the results under uniform or progressive rates makes sense.

Fig. \ref{ndes} shows the avalanche size distribution for systems of different sizes using the two rate forms. The curves display the typical form of a cut off power law: $N(S)\sim S^{-\tau}f(S/S_{max})$. No significant difference between uniform and progressive rates  is observed in the value of $\tau$, which is found to be $\tau\simeq 1.33$. The form of the cut off is rather well fitted by the function 
$f(S/S_{max})\sim \exp(-(S/S_{max})^{1.5})$. The fitted values of $S_{max}$ for different sizes $L$ are shown in the inset to Fig. \ref{ndes}. This plot defines the fractal dimension of the avalanches $d_f$, through $S_{max}\sim L^{d_f}$. The value obtained is $d_f\simeq 1.15$, independent again of the particular form of the rates used.

\begin{figure}
\includegraphics[width=8cm,clip=true]{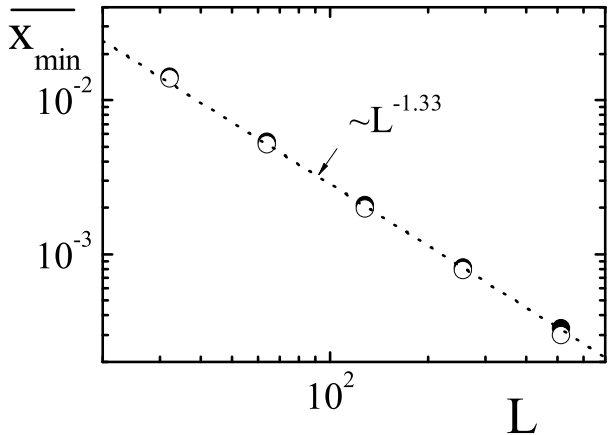}
\caption{Average strain increase for destabilization of avalanches, as a function of system size, for uniform (full) and progressive rate (empty symbols).
\label{theta}
}
\end{figure}

An important quantity in yielding transition theories is the exponent $\theta$ describing the distribution of shear transformation zones close to instability \cite{rosso_epl,pnas}.
It is defined by considering the distribution of distances to instability $x_i\equiv \sigma_i^{th}-\sigma_i$ 
of all sites in the system (in our case $\sigma_i^{th}\equiv 1$). In depinning transitions $\sigma_i$ 
can only increase as the dynamics proceeds, and this is why this exponent is trivial: $\theta_{(dep.)}=0$.
For yielding, however, the sign-alternating interaction kernel gives the possibility of non trivial $\theta$
values. In the mean field H\'ebraud-Lequeux approximation \cite{hl} $\theta=1$. In more realistic cases intermediate values 
are expected. The numerical determination of $\theta$ requires, from its definition, the evaluation of the histogram of
distances to instability, which can be done in principle for a single static configuration in the system in equilibrium (i.e., not while an avalanche is developing). 
There is also a second possibility for the evaluation of $\theta$ that considers the average value $\overline{x_{min}}$ of the stress increase between avalanches. It is found \cite{pnas} that $\overline {x_{min}}$ scales with system size as $\overline {x_{min}}\sim  L^{-d/(\theta+1)}$, allowing to determine $\theta$.

\begin{figure}
\includegraphics[width=8cm,clip=true]{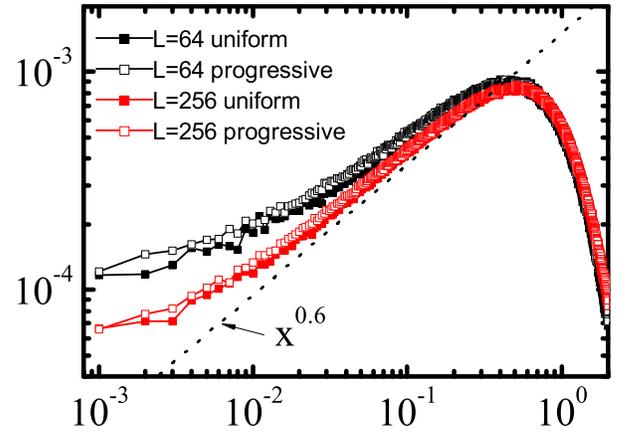}
\caption{Distribution of distances to instability of all sites in the system.
\label{theta2}
}
\end{figure}

I measured the average stress increase between avalanches for systems of different sizes 
and the results are contained in Fig. \ref{theta}.
The results for uniform and progressive rates are seen to be totally coincident, and the exponent is consistent with a value of $\theta\simeq 0.5$.
In addition, results based on the full distribution of distance to instabilities are contained in Fig. \ref{theta2}. As already found in \cite{pnas}, the values obtained with this method are a bit larger than the previous one. I find $\theta\simeq 0.6$, yet totally equivalent for the two different rate laws. 
It seems that finite size effects and corrections to scaling are still important for the system sizes analyzed, but the equivalence of results for uniform and progressive rates is clear.



Finally, I consider the second exponent for which a dependence on the rate law may be expected. This is the dynamical exponent $z$ that relates the avalanche size with the avalanche duration. The avalanche duration $T$ is expected to scale as a power law of the avalanche size $S$, namely $T\sim S^p$. The value of $p$ can be directly determined from the numerical simulations by fitting the $T$ vs $S$ correlation observed. The exponent $p$ is related to the dynamical exponent $z$ by $p=z/d_f$.

In the numerical determination of the duration $T$, the time needed to destabilize the first site in the avalanche was not considered. This was done in this way since for progressive rates, as the first site is destabilized by an infinitesimal quantity when increasing the applied stress quasistatically, it would add a diverging contribution to the total time, then completely spoiling the results (for uniform rates it only adds an additional time unit, but for consistency the first site was not considered either in this case).

\begin{figure}
\includegraphics[width=8cm,clip=true]{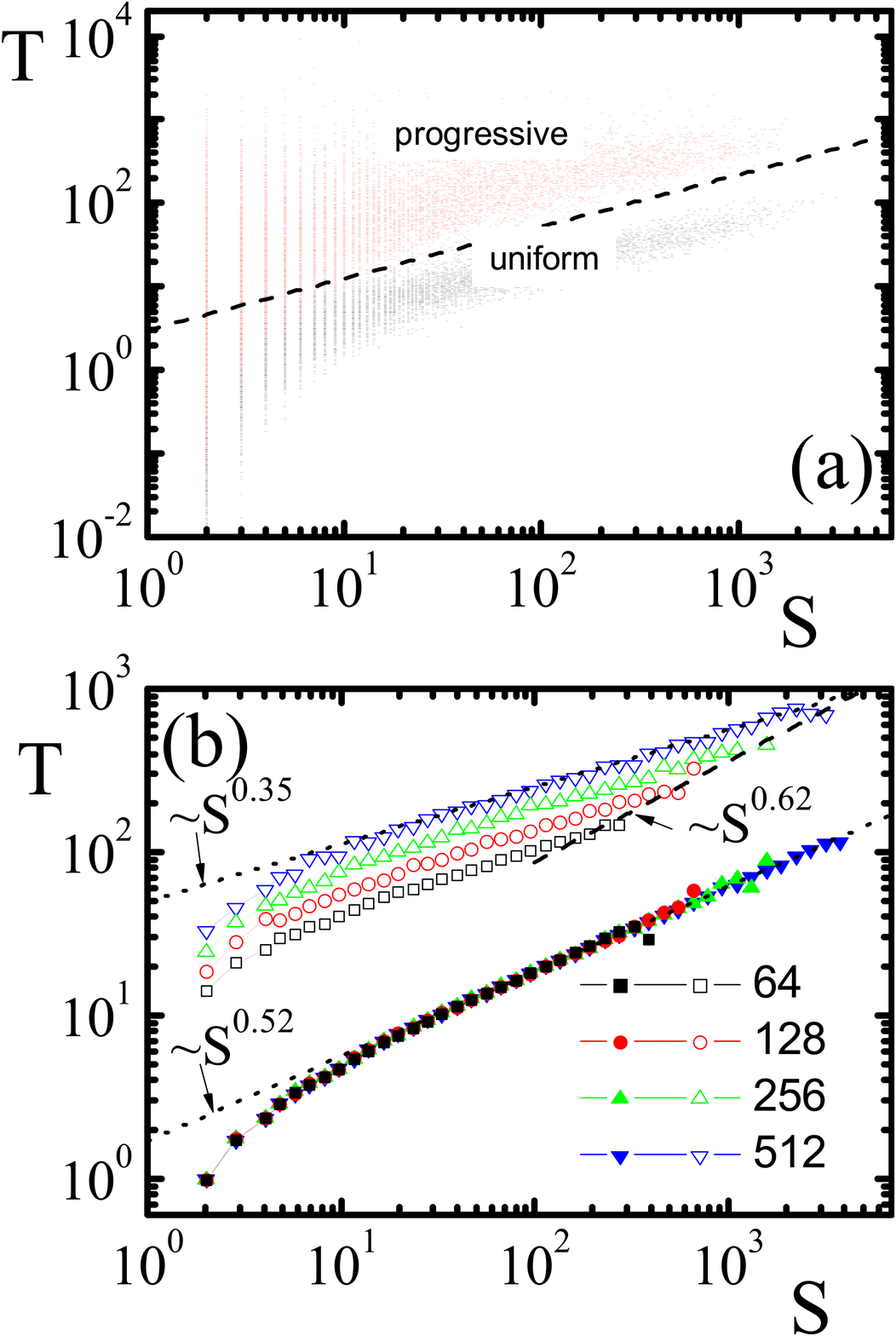}
\caption{(a)Duration-size plot of individual avalanches, in a system of size 512$\times$512. (b) Average duration as a function of avalanche size, for different system sizes, with uniform (full symbols) and progressive  rates (open symbols).
\label{zeta}
}
\end{figure}

A direct plot of $T$ vs $S$ from simulation in a system with $L=512$, using the two possible rate laws is presented in Fig. \ref{zeta}(a). From this kind of figures, and averaging the values of $T$ within small $S$ intervals, the curves in Fig. \ref{zeta}(b) are obtained. For uniform rates, the results show a consistent power law with an exponent $p\simeq 0.52$. Using the previously determined value $d_f\simeq 1.15$, it is obtained $z\simeq 0.6$. 

The results for progressive rates in Fig. \ref{zeta}(b) are definitely different from those for uniform rates. 
For a fixed system size it is obtained from the $T$ vs $S$ dependence that $p\simeq 0.35$ for the progressive case, providing $z=pd_f\simeq 0.4$, lower than the previous uniform case.
Yet, an additional behavior is observed. The results for progressively larger systems do not simply extend the region in which a power law is observed, but also produce a shift of the results towards larger values of $T$. 
This was an unexpected result, but it can be rationalized in the following way. Consider for simplicity the avalanches of size $S=2$. Since the first site is not taken into account, the duration of this avalanche is simply given by the activation time of the second site. Under uniform rates, this time is (on average) $T=\lambda=1$, and this is what is observed in Fig. \ref{zeta}(b). However, for progressive rates, the time will depend on the spatial distance $r$ from the position of the first site that is destabilized, to the second. As the strength $s$ of the kick scales as $1/r^2$, the time $T$ for destabilizing the second site (which coincides with the avalanche duration in this simple case) is expected to depend on $r$ as $T\sim 1/\sqrt s\sim r$, i.e, it increases with distance, and then with the possibility of larger values of $r$ as the system size increases.
This phenomenon allows to introduce an alternative definition of a dynamical exponent $\tilde z$, 
relating the maximum avalanche size with the maximum time duration, namely $T_{max}\sim S_{max}^{\tilde z/d_f}$. This dependence (sketched by the dotted line in Fig \ref{zeta}) provides in the present case $\tilde z/d_f\sim 0.62$. This previously unnoticed phenomenon is also present at a mean field level, where a more quantitative estimation of the value of $z$ can be given \cite{frj}.

\section{Further analysis and conclusions}

I summarize briefly the main results obtained. An elasto-plastic model was set up to study the plastic yielding transition of amorphous solid materials. Two variants on the model were considered, one that takes constant transition rates between different plastic states (the {\em uniform} case), and a second one (the {\em progressive} case) in which the transition rates are progressively larger as the local instability threshold is exceeded more and more.

Results for the critical exponents are contained in Table 1. 
It is found that for the uniform case the full set of critical exponents coincides with that reported in other implementations of EPM, in particular in Ref. \cite{pnas}. Some of the exponents for the progressive case (in concrete $\tau$, $d_f$, and $\theta$) coincide with those of the uniform case. Others ($\beta$ and $z$) are definitely different. In this last case an alternative definition of a dynamical exponent $\tilde z$ was given. Its value does not coincide either with the value of $z$ for the uniform case.

As the exponents obtained for the uniform case are similar to those obtained in Ref. \cite{pnas}, two scaling relations that were
proposed and verified there are also satisfied here, namely
\begin{equation}
\beta=1+\frac{z}{d-d_f}
\label{scaling1}
\end{equation}
and
\begin{equation}
\tau=2-\frac{\theta}{\theta+1}\frac{d}{d_f}
\label{scaling2}
\end{equation}

The progressive case was argued \cite{jagla2017} to apply to cases in which the plastic disorder potential is smooth, and this is the case that should correspond to most experimental situations. Static exponents in the progressive case coincide with those of the uniform case. This applies to $\tau$, $d_f$, $\theta$,  and also to the correlation length exponent $\nu$ that I did not measure directly, but that it must satisfy $\nu=1/(d-d_f)$, due to the statistical tilt symmetry of the problem\cite{sts}. The expected relation among critical exponents expressed by relation (\ref{scaling2}) is therefore satisfied in the progressive case also since it only involves static exponents.

\begin{table}
\begin{center}
\begin{tabular}{ |c|c|c| } 
 \hline
 & uniform & progressive \\ 
 &  rate &  rate \\ 
 \hline
$\beta$ &  $1.5\pm 0.1$ & $2.0\pm 0.1$ \\ 
$d_f$& $1.15\pm 0.05$ & $1.15\pm 0.05$ \\ 
$\theta$ & $0.55\pm 0.05$ & $0.55\pm 0.05$ \\ 
$\tau$  & $1.33\pm 0.03$ & $1.33\pm 0.03$ \\ 
$z/d_f$  & $0.52\pm 0.02$ & $0.35\pm 0.02$ \\ 
$\tilde z/d_f$  & undefined & $0.62\pm 0.05$ \\ 
\hline
\end{tabular}
\end{center}
\caption{Values of the critical exponents directly determined from the numerical simulations. Exponents $\beta$ and $z$ differ for uniform and progressive rate. Other exponents coincide.  }
\label{table:1}
\end{table}

Concerning the scaling relation (\ref{scaling1}), I note that $\beta$ is larger and $z$ is smaller in the progressive case compared with the uniform case. Since in the uniform case Eq. (\ref{scaling1}) is satisfied, it must be concluded that this equation is not satisfied for progressive rates. One may wonder if this relation is satisfied using the alternative exponent $\tilde z$, instead. The current numerical precision is not sufficient to give a definite answer to this question. In any case, it is clear that the theoretical argumentation leading to relation (\ref{scaling1}) must be reconsidered for the case of smooth disorder potentials.

I want to do an additional consideration about the values obtained for the $\beta$ exponent. Although it is tempting to think that these values are exactly given by $\beta=3/2$ (uniform) and $\beta=2$ (progressive), I emphasize that there is at present no strong argument supporting this claim. Yet, the analysis in Refs. \cite{jagla2017,prandtl_jagla} suggests that the difference between these two values should be exactly 1/2 in any dimension. An implementation of the present model in 3D in fact produces values of $\beta$ that adjust to this prediction, namely $\beta\simeq 1.35$ (3D uniform), and $\beta\simeq 1.85$ (3D progressive).



The dependence of some critical exponents on details of the transition probabilities is surprising in view of the universality that is expected in this kind of transition.
In this respect it must be noted that if the same protocol of uniform/progressive rates is applied to the depinning problem with short range interactions and for dimensions below the critical value, well defined values of $\beta$ and $z$ are obtained independently of the form of the rates. So the question may be posed: why the situation is different for yielding? I believe that the answer is that yielding is effectively a mean field transition because of the long range ($\sim r^{-d}$) form of the elastic interaction. In this limit, a dependence of $\beta$ and $z$ on the particular form of the rates is expected\cite{jagla2017} in the same way the these exponents also depend on the rates for fully connected depinning models\cite{fisher2}.

\end{document}